\documentclass[lettersize,journal]{IEEEtran}
\usepackage{amsmath,amsfonts,amssymb}
\usepackage{algorithmic}
\usepackage{algorithm}
\usepackage{array}
\usepackage[caption=false,font=normalsize,labelfont=sf,textfont=sf]{subfig}
\usepackage{textcomp}
\usepackage{stfloats}
\usepackage{url}
\usepackage{verbatim}
\usepackage{graphicx}
\usepackage{cite}
\usepackage{hyperref}
\usepackage{booktabs}
\usepackage{multirow}
\usepackage{array}
\newcolumntype{L}[1]{>{\raggedright\let\newline\\\arraybackslash\hspace{0pt}}m{#1}}
\newcolumntype{C}[1]{>{\centering\let\newline\\\arraybackslash\hspace{0pt}}m{#1}}
\newcolumntype{R}[1]{>{\raggedleft\let\newline\\\arraybackslash\hspace{0pt}}m{#1}}

\usepackage[dvipsnames]{xcolor}

\renewcommand{\paragraph}[1]{\noindent\textbf{#1}\ }

\begin{document}

\title{Bayesian Parameter-Efficient Fine-Tuning for Overcoming Catastrophic Forgetting}

\author{Haolin Chen, Philip N. Garner
\thanks{Haolin Chen is with the Idiap Research Institute, 1920 Martigny, Switzerland and \'Ecole Polytechnique F\'ed\'erale de Lausanne, 1015 Lausanne, Switzerland (email: haolin.chen@idiap.ch).}
\thanks{Philip N. Garner is with the Idiap Research Institute, 1920 Martigny, Switzerland (email: phil.garner@idiap.ch).}
}



\maketitle

\begin{abstract}
We are motivated primarily by the adaptation of text-to-speech synthesis models; however we argue that more generic parameter-efficient fine-tuning (PEFT) is an appropriate framework to do such adaptation. 
Nevertheless, catastrophic forgetting remains an issue with PEFT, damaging the pre-trained model's inherent capabilities. 
We demonstrate that existing Bayesian learning techniques can be applied to PEFT to prevent catastrophic forgetting as long as the parameter shift of the fine-tuned layers can be calculated differentiably. In a principled series of experiments on language modeling and speech synthesis tasks, we utilize established Laplace approximations, including diagonal and Kronecker-factored approaches, to regularize PEFT with the low-rank adaptation (LoRA) and compare their performance in pre-training knowledge preservation. Our results demonstrate that catastrophic forgetting can be overcome by our methods without degrading the fine-tuning performance, and using the Kronecker-factored approximation produces a better preservation of the pre-training knowledge than the diagonal ones.
\end{abstract}


\section{Introduction}
\IEEEPARstart{I}{n}
the context of text-to-speech synthesis (TTS), it has long been of interest to adapt a generic model to a specific domain such as a given speaker identity, language, or emotion.  The process is termed \emph{adaptation}; typically the generic model would be well-trained on a large dataset, whereas the (domain-specific) adaptation dataset would be too small to train a bespoke model.  Adaptation proved particularly useful in statistical parametric and neural TTS \cite{Yamagishi09,ArikCPPZ18}, and remains a goal of the recent Blizzard challenge \cite{perrotin23_blizzard}.
More recently, the state of the art in TTS is represented by more generic generative models that have arisen in the machine learning community, with advances made in the domains of text \cite{gpt3,gpt4}, vision \cite{ldm,imagen}, and audio \cite{audiolm,audiobox}, all feeding through to TTS.

A key paradigm that has emerged in the development and application of such generic models is the pre-training-fine-tuning approach, which involves initially training a model on a large dataset (pre-training) and subsequently fine-tuning it on a task-specific dataset. The paradigm has proven to be highly effective, leading to substantially more accurate and robust outcomes. More recent large pre-trained models have increasingly been equipped with in-context or zero-shot learning capabilities \cite{ldm,valle,audiobox}. However, when there are more data available for the target task, fine-tuning is still useful to further improve the performance considerably \cite{MosbachPRKE23}.
Notice that, whilst the vocabulary differs slightly, the goal is the same as for TTS.  It follows that current research in fine-tuning provides the means to adapt current TTS models.

The performance gains achieved by large pre-trained models are undeniably linked to their scale. Larger models, with their increased capacity, tend to deliver superior performance. However, as the size of pre-trained models increases, the costs associated with fine-tuning and storing all parameters become prohibitively high, making it practically infeasible. This has led to the study of parameter-efficient fine-tuning (PEFT) techniques \cite{HoulsbyGJMLGAG19,LiL20,bitfit,lora}, which optimize a small subset of the model parameters (either original parameters or additional ones) while leaving the rest unchanged, significantly reducing computation and storage costs. PEFT techniques have not only facilitated fine-tuning of large pre-trained models on low-resource devices but also enabled the easy sharing and deployment of customized models as far fewer parameters need to be stored and transferred.

Despite the benefits of (parameter-efficient) fine-tuning, it is not without its pitfalls. One significant risk is catastrophic forgetting \cite{MCCLOSKEY1989109,FRENCH1999128,GoodfellowMDCB13}, where the model loses much of the knowledge it gained during pre-training. This loss can adversely affect the model's ability to generalize to unseen data, a critical aspect of any machine learning model. The phenomenon is even more unfavorable on modern large pre-trained models that are usually multi-functional by training on a diverse range of tasks and data. For example, a language model may forget its general knowledge after continual instruction tuning \cite{Luo23}, or hypothetically, the controllability of emotions of a speech synthesizer may be compromised after fine-tuning on a specific voice.

Bayesian learning theory provides a principled solution to overcoming catastrophic forgetting. Considering optimizing the neural network as performing a maximum a posteriori (MAP) estimation of the network parameters given the fine-tuning data, it tries to find the optimal trade-off between the likelihood of the fine-tuning data and the prior knowledge of the pre-trained model, of which the latter is accessible in the form of the posterior over the parameters given the pre-training data. Although the true posterior is intractable, it can be approximated by fitting a Gaussian distribution with a mean equal to the MAP solution and a precision equal to the observed Fisher information. The technique is known as the Laplace approximation \cite{MacKay92a} and has been thoroughly studied \cite{EWC,KFAC,Botev17a,RitterBB18}.

In this paper, we demonstrate quite generally that existing Bayesian learning techniques can be applied to PEFT to overcome catastrophic forgetting. Deriving from the Bayesian transfer learning framework, we show that it is viable to regularize the PEFT to preserve the pre-training knowledge as long as the parameter shift of the fine-tuned layers can be expressed in a differentiable manner. Utilizing established Laplace approximation techniques including diagonal \cite{EWC,L2SP} and Kronecker-factored \cite{KFAC,KFAC_CL} approximations of the Hessian, we conduct a series of experiments on language modeling and speech synthesis tasks with the low-rank adaptation (LoRA) \cite{lora} to demonstrate the effectiveness and compare the performance of different methods. Specifically, we start from a study on text classification and causal language modeling tasks, the quantitative nature of which allows both rigorous comparison of techniques and comparison with existing literature. We then verify our findings on our target task of speaker adaptation of speech synthesis, where the results are typically more subjective and more onerous to generate. Our results demonstrate that catastrophic forgetting can be overcome by such methods without degrading the fine-tuning performance, and the Kronecker-factored approximations generate a better preservation of the pre-training knowledge than the diagonal ones. Audio samples and source code are available\footnote{\url{https://github.com/idiap/bayesian-peft}}.

\section{Related Work}
\subsection{Laplace Approximation}
The Laplace approximation \cite{MacKay92a} is an established technique in statistics and machine learning to approximate a complex posterior distribution with a Gaussian distribution. This is achieved by identifying the mode of the posterior distribution, which is the maximum a posteriori estimate, and then approximating the distribution around this mode using a second-order Taylor expansion. Two popular kinds of Laplace approximation are the diagonal approximation \cite{EWC,L2SP}, which only considers the variance of each parameter itself and ignores the interactions between parameters, and the Kronecker-factored approximation \cite{KFAC} that also takes the covariance between parameters within each layer into account. Thanks to the additional information on the off-diagonal elements of the Hessian, the Kronecker-factored approximation has been shown to be more accurate than the diagonal approximation in capturing the loss landscape \cite{KFAC_CL}.

The Laplace approximation has been widely applied in neural network optimization (natural gradient descent) \cite{Pascanu13,KFAC,Botev17a,EKFAC}, improving calibration of neural networks (predictive uncertainty estimation) \cite{RitterBB18,Kristiadi0H20,ImmerKB21,DaxbergerKIEBH21}, and overcoming catastrophic forgetting in transfer and continual learning \cite{EWC,KFAC_CL,ncl}. In this work, we focus on its application in mitigating catastrophic forgetting in the PEFT setting.

\subsection{Parameter-Efficient Fine-Tuning}
There exists a variety of PEFT techniques taking different approaches to adding new trainable components to, or modifying existing parameters of the pre-trained model. Representative PEFT techniques include
\begin{enumerate}
\item inserting serial or parallel adapters with a bottleneck structure to the model \cite{HoulsbyGJMLGAG19,PfeifferVGR20,HeZMBN22},
\item prepending trainable tokens to the input and hidden states of the transformer block \cite{LiL20,LesterAC21},
\item fine-tuning the bias terms inside the model only \cite{bitfit},
\item optimizing the low-rank approximation of the change of weights \cite{lora,loha,lokr,LyCORIS}, and
\item the combination of the above methods \cite{HeZMBN22,MaoMHAM0YK22}. 
\end{enumerate}

\subsection{Continual Learning}
Continual learning aims to enable the model to learn from non-stationary streams of data.
\cite{VenTT22} categorizes continual learning into three types: task-, domain-, and class-incremental learning. 
In the context of the adaptation of TTS models, we are interested in the scenario where the pre-trained model is fine-tuned to solve the same task as the pre-training one using data from different domains. This is an example of the domain-incremental type. 
Despite close ties with continual learning, the scenario concerned aligns better with \textit{transfer learning} and \textit{domain adaptation}. 
Further constraints that should be considered include that not all pre-training data are accessible and that the pre-training process cannot be replayed.
All such constraints limit the usage of techniques designed for task- and class-incremental learning, such as Learning without Forgetting \cite{LiH16} and Synaptic Intelligence \cite{ZenkePG17}.

There have been attempts to utilize PEFT techniques, mainly the low-rank adaptation (LoRA), in the continual learning setting. C-LoRA \cite{c-lora} leverages a self-regularization mechanism with LoRA to prevent catastrophic forgetting in continual customization of text-to-image models; O-LoRA \cite{o-lora} continually learns tasks in different low-rank subspaces that are kept orthogonal to each other to minimize interference. For general fine-tuning, \cite{ewclora} proposes to regularize the LoRA weights with Elastic Weight Consolidation \cite{EWC} when fine-tuning language models on question-answering tasks while preserving their general inference abilities.

\section{Bayesian Transfer Learning}
\subsection{Framework}
The optimization of neural networks can be interpreted as performing a maximum a posteriori (MAP) estimation of the network parameters $\theta$ given the training data. In the transfer learning setting, the model has been pre-trained on a task $\mathcal{A}$ using data $\mathcal{D}_\mathcal{A}$, and is then fine-tuned on a downstream task $\mathcal{B}$ using data $\mathcal{D}_\mathcal{B}$. The overall objective is to find the optimal parameters on task $\mathcal{B}$ while preserving the prior knowledge of the pre-trained model on task $\mathcal{A}$. The posterior to be maximized in the MAP estimation can be written as:
\begin{equation} \label{eq:map_posterior}
    \begin{aligned}
        p(\theta|\mathcal{D}_\mathcal{A},\mathcal{D}_\mathcal{B})&=\frac{p(\mathcal{D}_\mathcal{B}|\theta,\mathcal{D}_\mathcal{A})p(\theta|\mathcal{D}_\mathcal{A})}{p(\mathcal{D}_\mathcal{B}|\mathcal{D}_\mathcal{A})} \\
        &=\frac{p(\mathcal{D}_\mathcal{B}|\theta)p(\theta|\mathcal{D}_\mathcal{A})}{p(\mathcal{D}_\mathcal{B})}
    \end{aligned}
\end{equation}
where $\mathcal{D}_\mathcal{B}$ is assumed to be independent of $\mathcal{D}_\mathcal{A}$. Taking a logarithm of the posterior, the MAP objective is therefore:
\begin{equation} \label{eq:map_objective}
    \begin{aligned}
        \theta^*&=\underset{\theta}{\arg\max}\,\log p(\theta|\mathcal{D}_\mathcal{A},\mathcal{D}_\mathcal{B}) \\
        &=\underset{\theta}{\arg\max}\,[\log p(\mathcal{D}_\mathcal{B}|\theta)+\log p(\theta|\mathcal{D}_\mathcal{A})]
    \end{aligned}
\end{equation}

The first term $p(\mathcal{D}_\mathcal{B}|\theta)$ is the likelihood of the data $\mathcal{D}_\mathcal{B}$ given the parameters $\theta$, which can be expressed as the training loss function on task $\mathcal{B}$, denoted by $\mathcal{L}_\mathcal{B}(\theta)$. The second term $p(\theta|\mathcal{D}_\mathcal{A})$ is the posterior of the parameters given the pre-training data $\mathcal{D}_\mathcal{A}$. If training the network from scratch, i.e., assuming $\mathcal{D}_\mathcal{A}$ and $\mathcal{D}_\mathcal{B}$ to be one dataset $\mathcal{D}$, this term is usually approximated by a zero-mean isotropic Gaussian distribution, i.e., $p(\theta|\mathcal{D})=\mathcal{N}(\theta|0,\sigma^2\mathbf{I})$, corresponding to the $\mathcal{L}_2$ regularization. However, for transfer learning, this posterior must encompass the prior knowledge of the pre-trained model to reflect which parameters are important for task $\mathcal{A}$. Despite the true posterior being intractable, $\log p(\theta|\mathcal{D}_\mathcal{A})$ can be defined as a function $f(\theta)$ and approximated around the optimum point $f(\theta_0)$ \cite{MacKay92a}, where $\theta_0$ is the pre-trained values and $\nabla f(\theta_0) = 0$. Performing a second-order Taylor expansion on $f(\theta)$ around $\theta_0$ gives:
\begin{equation} \label{eq:posterior_approx}
    \begin{aligned}
        \log p(\theta|\mathcal{D}_\mathcal{A})&\approx f(\theta_0)+\frac{1}{2}(\theta-\theta_0)^\top\nabla^2f(\theta_0)(\theta-\theta_0) \\
        &=f(\theta_0)+\frac{1}{2}(\theta-\theta_0)^\top\mathbf{H}(\theta-\theta_0)
    \end{aligned}
\end{equation}
where $\mathbf{H}$ is the Hessian matrix of $f(\theta)$ at $\theta_0$. The second term suggests that the posterior of the parameters on the pre-training data can be approximated by a Gaussian distribution with mean $\theta_0$ and covariance $\mathbf{H}^{-1}$. Note that the negation of the expected value of the Hessian over the data distribution is the Fisher information matrix (FIM) $\mathbf{F}$, i.e., $\mathbf{F}=-\mathbb{E}_{\mathcal{D}_\mathcal{A}}[\mathbf{H}]$. Following Equation \ref{eq:map_objective}, the training objective becomes:
\begin{equation} \label{eq:map_obj_final}
    \begin{aligned}
        \theta^*&=\underset{\theta}{\arg\min}\,[\mathcal{L}_\mathcal{B}(\theta)-\frac{1}{2}(\theta-\theta_0)^\top\mathbf{H}(\theta-\theta_0)]
    \end{aligned}  
\end{equation}

Finally, the loss function that we minimize during fine-tuning can be written as:
\begin{equation} \label{eq:map_loss}
    \begin{aligned}
        \mathcal{L}(\theta)&=\mathcal{L}_\mathcal{B}(\theta)+\lambda(\theta-\theta_0)^\top\mathbf{F}(\theta-\theta_0)
    \end{aligned}
\end{equation}
where $\lambda$ is the regularization strength that determines how much prior knowledge should be preserved during fine-tuning. 

\subsection{Diagonal Approximation of the Hessian}
Modern neural networks typically have millions to billions of parameters, thus the Hessian, being at least terabytes, is intractable to compute and store. One practical approximation of the Hessian is the diagonal of the Fisher information matrix, i.e., the expected square of the gradients over the data distribution, known as Elastic Weight Consolidation (EWC) \cite{EWC}. The loss function of EWC is:
\begin{equation} \label{eq:ewc_loss}
    \begin{aligned}
        \mathcal{L}_{EWC}(\theta)&=\mathcal{L}_\mathcal{B}(\theta)+\lambda\mathbf{F}_{EWC}(\theta-\theta_{0})^2
    \end{aligned}
\end{equation}
where $\mathbf{F}_{EWC}$ is the vectorized expected square of the gradients over the distribution of $\mathcal{D}_\mathcal{A}$. 

To estimate $\mathbf{F}_{EWC}$, a small subset of the pre-training data $\mathcal{D}_\mathcal{A}$ is sampled and used to compute the gradients of the training loss function $\mathcal{L}_\mathcal{A}(\theta)$ on task $\mathcal{A}$. The final $\mathbf{F}_{EWC}$ is then the average of the square gradients over the sampled data. 

A simplified version of EWC, named L2-SP \cite{L2SP}, assigns equal importance to all parameters, which is equivalent to assuming that the Fisher information matrix is an identity matrix. The loss function of L2-SP is:
\begin{equation} \label{eq:l2sp_loss}
    \begin{aligned}
        \mathcal{L}_{L2-SP}(\theta)&=\mathcal{L}_\mathcal{B}(\theta)+\lambda(\theta-\theta_{0})^2
    \end{aligned}
\end{equation}

L2-SP can be regarded as an extension of the $\mathcal{L}_2$ regularization: instead of zero, it limits the parameters to be close to the pre-trained values during fine-tuning by assigning a Gaussian prior $\mathcal{N}(\theta_0,\sigma^2\mathbf{I})$. Despite being overly simplified, L2-SP proves to be effective in preventing catastrophic forgetting in transfer learning \cite{L2SP}, and is particularly useful when the pre-training data are unavailable since no estimation of the FIM is required.

\subsection{Kronecker-Factored Approximation of the Hessian}
While first-order approximations such as EWC and L2-SP are simple and efficient, they are not accurate enough to capture the complete loss landscape since they ignore the off-diagonal elements of the Hessian, i.e., the interactions between parameters. To address this issue, recent advances in second-order optimization \cite{KFAC,Botev17a} utilize block-diagonal approximations of the Hessian: the diagonal blocks of the Hessian, corresponding to the interactions between parameters within a single layer, can be approximated as a Kronecker product of two much smaller matrices. This approximation is known as the Kronecker-factored approximate curvature, usually abbreviated as KFAC.

Following \cite{KFAC}, we denote the input, the weight, the pre-activations, the non-linear function, and the output of the $l$-th layer as $a_{l-1}$, $W_l$, $s_l$, $\phi_l$ and $a_l$, respectively. For simplicity, we only consider linear layers with no bias term, thus $s_l = W_l a_{l-1}$ and $a_l = \phi_l(s_l)$. We further define $g_l = \frac{\partial \mathcal{L}}{\partial s_l}$ as the gradient of the loss function $\mathcal{L}$ with respect to the pre-activations $s_l$. The FIM with respect to the weights $W_l$ can be written as:
\begin{equation} \label{eq:kfac_fim}
    \begin{aligned}
        \mathbf{F}_{KFAC}^l&=\frac{\partial^2 \mathcal{L}}{\partial^2 \mathtt{vec}(W_l)} = A_l \otimes G_l
    \end{aligned}
\end{equation}
where $\mathtt{vec}(W_l)$ is the vectorized form of $W_l$, $A_l = a_{l-1} a_{l-1}^\top$, $G_l = g_l g_l^\top$ and $\otimes$ is the Kronecker product operator. To calculate the expectation, the two factors are assumed to be independent, thus the expected Kronecker product is approximated as the Kronecker product of the expected factors. Thanks to a property of the Kronecker product, the quadratic penalty term for each layer can be efficiently calculated:
\begin{equation} \label{eq:kfac_feature}
    \begin{aligned}
        (A_l \otimes G_l) \mathtt{vec}(\Delta W_l) = \mathtt{vec}(G_l \Delta W_l A_l)
    \end{aligned}
\end{equation}
where $\Delta W_l = W_l - W_l^0$ is the parameter shift from the pre-trained weight $W_l^0$ of the $l$-th layer. The overall loss function of KFAC is:
\begin{equation} \label{eq:kfac_loss}
    \begin{aligned}
        \mathcal{L}_{KFAC}(\theta)=&\mathcal{L}_\mathcal{B}(\theta) + \\
        &\lambda\sum_{l=1}^L \mathtt{vec}(\Delta W_l) * \mathtt{vec}(G_l \Delta W_l A_l)
    \end{aligned}
\end{equation}

Despite KFAC's assumption of independence between layers, the most important in-layer parameter interactions are taken into account. It has been demonstrated that KFAC leads to better prior knowledge preservation in continual learning than using a diagonal approximation of the Hessian \cite{KFAC_CL}.

\section{Bayesian PEFT}
In this work, we aim to show that Bayesian transfer learning can provide a unifying framework for a variety of PEFT techniques.  Such an approach not only retains the parameter efficiency of PEFT but also brings a principled approach to regularization, in turn overcoming catastrophic forgetting.

Looking back on Eq. \ref{eq:map_loss}, it is not difficult to see that, as long as the parameter shift $\Delta W_l$ of the fine-tuned layers can be expressed in a differentiable way, the Bayesian transfer learning framework can be applied to any PEFT technique in the form of modification to the inherent weight of the pre-trained model. The loss function of Bayesian transfer learning with PEFT is therefore:
\begin{equation} \label{eq:peft_loss}
    \begin{aligned}
        \mathcal{L}_{PEFT}(\theta)=&\mathcal{L}_\mathcal{B}(\theta) + \\
        &\lambda\sum_{l=1}^L\mathtt{vec}(\Delta W_l)^\top\mathbf{F}_l\mathtt{vec}(\Delta W_l)
    \end{aligned}
\end{equation}

The most representative PEFT technique that fits this requirement is the low-rank adaptation (LoRA) family. LoRA \cite{lora} aims to optimize the low-rank approximation of the change of the original weight matrices based on the hypothesis that the change of weights during fine-tuning has a low intrinsic rank. It is formulated as adding the matrix product of two low-rank matrices to the original weight matrix, i.e., $W_l = W_l^0 + \gamma A_l B_l^\top$, where $W_l^0 \in \mathbb{R}^{d_o \times d_i}$ is the pre-trained weight matrix, $\gamma$ is a scaling factor, $A_l \in \mathbb{R}^{d_o \times r}$ and $B_l \in \mathbb{R}^{d_i \times r}$ are two low-rank matrices. Therefore, the weight modification (delta weight) of each layer is simply $\Delta W_l = \gamma A_l B_l^\top$. Following Eq. \ref{eq:peft_loss}, the loss function of Bayesian transfer learning with LoRA is:
\begin{equation} \label{eq:lora_loss}
    \begin{aligned}
        \mathcal{L}_{LoRA}(\theta)=&\mathcal{L}_\mathcal{B}(\theta) + \\
        &\lambda\sum_{l=1}^L \mathtt{vec}(\gamma A_l B_l^\top)^\top \mathbf{F}_l \mathtt{vec}(\gamma A_l B_l^\top)
    \end{aligned}
\end{equation}

Apart from the original LoRA, there exist several variants of LoRA including AdaLoRA \cite{adalora}, which adaptively assigns the rank to the LoRA matrices in each layer, FedPara (LoHa) \cite{loha,LyCORIS}, of which the delta weight is the Hadamard product of two LoRA delta weights, and KronA (LoKr) \cite{lokr,LyCORIS}, which generates the delta weight by the Kronecker product of two low-rank matrices. Thanks to the explicit formulation of the delta weight, the LoRA family fits any aforementioned approximation of the Hessian in the Bayesian transfer learning framework. 
We also note that other PEFT methods such as $(\mathtt{IA})^3$ \cite{ia3} and Orthogonal Butterfly \cite{boft}, that do not explicitly calculate the delta weight, also fit in the framework, although regularizing these methods may require extra computation and memory. Given that the original LoRA has achieved sufficiently good performance, e.g., it matches the full fine-tuning performance on the GLUE benchmark \cite{lora}, and other LoRA variants only offer insubstantial improvements, we only employ the original LoRA and focus on the study of regularization methods in our experiments.

\section{Experiments: Language Modeling}
\subsection{Tasks}
We first apply our methods to fine-tuning pre-trained language models with LoRA on two sets of language modeling tasks: text classification and causal language modeling. The reason for this choice of task is twofold: The first is that language models can be evaluated quantitatively; a clear metric is associated with each task. The second is that it allows objective comparison with the wider literature.

\subsubsection{Text Classification}
We select three sentence-pair classification tasks and one single-sentence classification task from the GLUE benchmark \cite{glue}. The sentence-pair tasks are: MNLI \cite{mnli}, a natural language inference task of predicting whether a premise entails, contradicts or is neutral to a hypothesis, QQP \cite{qqp}, a paraphrase detection task of predicting whether a pair of sentences are semantically equivalent, and QNLI \cite{qnli}, a question answering task of predicting whether a sentence answers a question. The single-sentence task is SST-2 \cite{sst2}, a sentiment analysis task of predicting whether a sentence has positive or negative sentiment. For all tasks, the fine-tuning performance is reflected by the accuracy on the validation set. The number of training examples in the four selected datasets are MNLI: 393k, QQP: 363k, QNLI: 105k, and SST-2: 67k.

\subsubsection{Causal Language Modeling}
We experiment on the two subsets, WikiText-2 and WikiText-103, of the WikiText dataset \cite{wikitext}, a collection of over 100 million tokens extracted from the set of verified good and featured articles on Wikipedia. The number of tokens in WikiText-2 and WikiText-103 are 2.1M and 103M, respectively. The fine-tuning performance is reflected by the perplexity on the validation set, which is shared by the two subsets.

\subsection{Model: OPT}
We select the Open Pre-trained Transformers (OPTs) \cite{opt} with 350M and 1.3B parameters as the pre-trained models for our experiments. The OPTs are a suite of decoder-only transformers ranging from 125M to 175B parameters pre-trained on a series of large open-access corpora, including a subset of the Pile \cite{pile}. Our choice of model sizes is based on those of state-of-the-art pre-trained TTS models ranging from 100M to 1B parameters \cite{styletts2,audiobox,basetts}, so that the findings will hopefully provide useful guidance for our target task.

For text classification, a classification head is added on the last token the model generates and trained along with LoRA. This is purely for the simplicity of the implementation, though it could also be done by instruction tuning. For causal language modeling, the model structure remains unchanged.

\begin{table*}[htbp]
    \centering
    \caption{Main results of language modeling experiments.}
      \begin{tabular}{clcccccccc}
      \toprule
      \multirow{2}[4]{*}{\textbf{Model}\vspace{0.75em}} & \multicolumn{1}{c}{\multirow{2}[4]{*}{\textbf{Method}\vspace{0.75em}}} & \multirow{2}[4]{*}{\textbf{$\lambda$}\vspace{0.75em}} & \multirow{2}[4]{*}{\textbf{PT PPL}\vspace{0.75em}} & \multicolumn{4}{c}{\textbf{Classification (ACC$\uparrow$\,/\,PPL$\downarrow$)}} & \multicolumn{2}{c}{\textbf{CLM (PPL$\downarrow$\,/\,PPL$\downarrow$)}} \\
  \cmidrule{5-10}          &       &       &       & \textbf{MNLI} & \textbf{QQP} & \textbf{QNLI} & \textbf{SST-2} & \textbf{WikiText-2} & \textbf{WikiText-103} \\
      \midrule
      \multirow{5}[2]{*}{OPT-350M\vspace{1.5em}} & None  & -     & \multirow{5}[2]{*}{15.40\vspace{1.5em} } & 83.33\%\,/\,523.7 & 88.97\%\,/\,1234\kern 0.16667em & 89.79\%\,/\,51.11 & 93.81\%\,/\,19.05 & 13.48\,/\,20.35 & 15.21\,/\,31.74 \\
            & L2-SP & $10^{-3}$ &       & 83.35\%\,/\,33.65 & 88.28\%\,/\,19.91 & 89.84\%\,/\,23.69 & 93.72\%\,/\,16.66 & 13.62\,/\,18.21 & 15.95\,/\,20.61 \\
            & EWC   & $10^{4}$ &       & 83.67\%\,/\,18.67 & 88.73\%\,/\,15.94 & 89.88\%\,/\,16.91 & 93.78\%\,/\,15.60 & 13.55\,/\,17.17 & 15.80\,/\,16.87 \\
            & KFAC  & $10^{6}$ &       & 84.21\%\,/\,17.24 & 89.28\%\,/\,15.80 & 90.13\%\,/\,16.41 & 93.76\%\,/\,15.56 & 13.59\,/\,16.22 & 15.60\,/\,16.08 \\
      \midrule
      \multirow{5}[2]{*}{OPT-1.3B\vspace{1.5em}} & None  & -     & \multirow{5}[2]{*}{11.18\vspace{1.5em} } & 87.70\%\,/\,23.55 & 90.97\%\,/\,16.28 & 92.59\%\,/\,13.45 & 95.94\%\,/\,11.87 & 9.81\,/\,13.08 & 10.53\,/\,24.32 \\
            & L2-SP & $10^{-4}$ &      & 87.77\%\,/\,15.66 & 90.32\%\,/\,15.94 & 92.51\%\,/\,13.33 & 96.10\%\,/\,11.78 & 9.82\,/\,12.72 & 10.71\,/\,15.93 \\
            & EWC   & $10^{4}$ &       & 87.78\%\,/\,11.72 & 90.62\%\,/\,11.32 & 92.41\%\,/\,11.40 & 96.08\%\,/\,11.23 & 9.81\,/\,11.89 & 10.70\,/\,13.45 \\
            & KFAC  & $10^{5}$ &       & 87.76\%\,/\,11.45 & 90.64\%\,/\,11.25 & 92.28\%\,/\,11.43 & 96.17\%\,/\,11.20 & 9.84\,/\,11.73 & 10.70\,/\,11.55 \\
      \bottomrule
      \end{tabular}%

    \medskip
    \raggedright
    \hspace{0.5em}* ACC: accuracy, PPL: perplexity, PT PPL: perplexity of pre-trained model on the sampled test set from the Pile, CLM: causal language modeling.

    \label{tab:lm_main_results}%
\end{table*}%

\subsection{Experimental Details}
\paragraph{Implementation.} We base our code on the text classification and the causal language modeling examples of the Hugging Face Transformers library \cite{hf-transformers}. The Bayesian transfer learning techniques are implemented with the Hugging Face Parameter-Efficient Fine-Tuning (PEFT) library \cite{hf-peft}.

\paragraph{Hessian estimation.} The Hessian estimates are computed on the pre-training task, i.e., the causal language modeling task, and are shared by all fine-tuning tasks. We randomly sample 20,000 examples from the subset of the Pile used to pre-train the OPTs to compute the Hessian estimates for EWC and KFAC, and another 2,000 examples for the evaluation of the pre-training knowledge preservation.

\paragraph{Training and evaluation.} All models are trained using the Adam optimizer \cite{adam} on each dataset for 3 epochs without weight decay. The learning rate is set to $5 \times 10^{-4}$ for the 350M model and $2\times 10^{-4}$ for the 1.3B model, both with a linear decay schedule. For the text classification tasks, the batch size for all models is set to 32, while for the causal language modeling tasks, the batch size is set to 16 for the 350M model and 8 for the 1.3B model with a context window of 1024 tokens. LoRA is applied to the linear modules that produce the query and value in every self-attention module. The rank and the scaling factor of LoRA are set to 16 and 2 respectively for all models, resulting in the percentage of trainable parameters of the 350M and 1.3B model being 0.473\% and 0.239\%, respectively. 
To evaluate the fine-tuning performance, we calculate the accuracy or the perplexity on the validation set for the text classification tasks and the causal language modeling tasks respectively. For MNLI, the ``matched'' validation set is used.
For the evaluation of the pre-training knowledge preservation, we calculate the perplexity on the sampled test set of the Pile. 
We run a coarse hyper-parameter sweep on the regularization strength $\lambda$ with a step size of 10 times for each method on each task. The optimal $\lambda$ is selected balancing the fine-tuning performance and the preservation of pre-training knowledge, typically the point where fine-tuning performance is going to drop greatly if the regularization further strengthens. 
All experiments were conducted on machines equipped with one NVIDIA RTX3090. The results are averaged over 5 runs with different random seeds.

\subsection{Results and Analyses}
The main results are shown in Table \ref{tab:lm_main_results}. Note that the method ``None'' refers to LoRA without regularization. We elaborate our findings from several perspectives.

\paragraph{Catastrophic forgetting.} Compared to the pre-trained models, all models fine-tuned without regularization demonstrated significant forgetting of the pre-training knowledge, e.g., the perplexity on the pre-training data increased from 15.40 to 523.7 when fine-tuned on MNLI. Comparing different tasks, it is obvious that the forgetting is more severe when the model is fine-tuned on more data. In terms of model sizes, we notice that larger models tend to forget the pre-training knowledge less than smaller models, which suggests larger models have better resistance to catastrophic forgetting.

\paragraph{Comparison of regularization methods.} All regularization methods significantly reduced the loss of pre-training knowledge. Among them, L2-SP underperforms other methods by a large margin, which is reasonable given its over-simplified assumption of diagonal Hessian with equal importance on all parameters. In general, the Kronecker-based methods outperform EWC especially when there is more fine-tuning data, however, the difference is less significant for larger models. This demonstrates that knowledge preservation does benefit from more accurate Hessian estimations.

\begin{table}[htbp]
    \centering
    \caption{Comparison of performance with varying regularization strength of OPT-350M on MNLI.}
      \begin{tabular}{lccc}
      \toprule
      \textbf{Method} & \textbf{$\lambda$} & \textbf{Accuracy$\uparrow$} & \textbf{Perplexity$\downarrow$} \\
      \midrule
      Pre-trained & -     & -     & 15.40  \\
      \midrule
      None  & -     & 83.33\% & 523.74  \\
      \midrule
      \multirow{3}[2]{*}{L2-SP\vspace{0.5em}} & $10^{-4}$ & 84.52\% & 52.51  \\
            & $\mathbf{10^{-3}}$ & \textbf{83.35\%} & \textbf{33.65} \\
            & $10^{-2}$ & 81.51\% & 34.23  \\
      \midrule
      \multirow{3}[2]{*}{EWC\vspace{0.5em}} & $10^{3}$ & 84.11\% & 26.84  \\
            & $\mathbf{10^{4}}$ & \textbf{83.67\%} & \textbf{18.67} \\
            & $10^{5}$ & 82.03\% & 16.88  \\
      \midrule
      \multirow{3}[2]{*}{KFAC\vspace{0.5em}} & $10^{5}$ & 84.32\% & 19.38  \\
            & $\mathbf{10^{6}}$ & \textbf{84.21\%} & \textbf{17.24} \\
            & $10^{7}$ & 83.12\% & 17.10  \\
      \bottomrule
      \end{tabular}%
    \label{tab:lm_lambda}%
  \end{table}%

\paragraph{Regularization strength.} We provide an example of the regularization strength $\lambda$ sweep for the 350M model fine-tuned on MNLI, which is shown in Table \ref{tab:lm_lambda}. As $\lambda$ increases, the parameters are more constrained to the pre-trained values, thus the fine-tuning performance drops. We select the optimal $\lambda$ as the one that achieves a fine-tuning performance better than that of using the original LoRA and has the lowest perplexity on the pre-training data. It can be seen that, compared to KFAC-based methods, the pre-training knowledge preservation of EWC is worse when achieving the same level of fine-tuning performance. We also observe that the fine-tuning benefits from the regularization when $\lambda$ is small, which can be attributed to the fact that the Hessian estimation introduces a Gaussian prior that better describes the loss landscape than assuming an isotropic Gaussian prior at zero. This suggests that Bayesian transfer learning can lead to better fine-tuning performance as well as overcoming catastrophic forgetting. 

\begin{table}[htbp]
    \centering
    \caption{Comparison of Hessian estimates with varying samples.}
    \resizebox{\columnwidth}{!}{
      \begin{tabular}{clcccc}
      \toprule
      \multirow{2}[4]{*}{\textbf{Model}\vspace{0.75em}} & \multicolumn{1}{c}{\multirow{2}[4]{*}{\textbf{Samples}\vspace{0.75em}}} & \multicolumn{2}{c}{\textbf{MNLI}} & \multicolumn{2}{c}{\textbf{WikiText-103}} \\
  \cmidrule{3-6}          &       & \textbf{EWC} & \textbf{KFAC} & \textbf{EWC} & \textbf{KFAC} \\
      \midrule
      \multirow{4}[2]{*}{OPT-350M\vspace{0.5em}} & 20000 & 83.67\%\,/\,18.67 & 84.21\%\,/\,17.24 & 15.80\,/\,16.87 & 15.60\,/\,16.08 \\
            & 2000  & 83.66\%\,/\,18.77 & 84.30\%\,/\,17.64 & 15.80\,/\,16.96 & 15.57\,/\,16.22 \\
            & 200   & 83.71\%\,/\,18.50 & 84.51\%\,/\,17.60 & 15.83\,/\,16.84 & 15.47\,/\,16.79 \\
            & 20    & 83.59\%\,/\,18.63 & 84.47\%\,/\,21.39 & 15.83\,/\,16.96 & 15.37\,/\,18.50 \\
      \midrule
      \multirow{4}[2]{*}{OPT-1.3B\vspace{0.5em}} & 20000 & 87.78\%\,/\,11.72 & 87.76\%\,/\,11.45 & 10.70\,/\,13.45 & 10.70\,/\,11.55 \\
            & 2000  & 87.79\%\,/\,11.74 & 87.70\%\,/\,11.46 & 10.70\,/\,13.36 & 10.70\,/\,11.53 \\
            & 200   & 87.74\%\,/\,11.70 & 87.76\%\,/\,11.54 & 10.71\,/\,13.22 & 10.66\,/\,11.68 \\
            & 20    & 87.85\%\,/\,11.67 & 87.71\%\,/\,11.94 & 10.70\,/\,13.49 & 10.59\,/\,12.53 \\
      \bottomrule
      \end{tabular}%
    }
    \label{tab:lm_vary_data}%
\end{table}%

\paragraph{Hessian estimates with varying samples.} We further experiment on Hessian estimates with a reduced amount of pre-training data to investigate the effect of the sample size on the accuracy of the Hessian estimation. The results are shown in Table \ref{tab:lm_vary_data}. We observe that EWC is more robust to the sample size than KFAC, showing no degradation in pre-training knowledge preservation with Hessian estimates on fewer samples, whereas KFAC demonstrates significant degradation in perplexity on the pre-training data when the sample size is reduced to 20. This can also be corroborated by the increasing fine-tuning performance of KFAC when sample sizes decrease, which signifies less effective regularization. However, for other larger sample sizes, KFAC always outperforms EWC. Overall, the results suggest that KFAC, while being superior to EWC, requires more data to be estimated accurately than EWC, which is reasonable given its additional off-diagonal elements in the Hessian estimation.

\paragraph{Computational cost and memory usage.}
We compare the computational cost and memory usage of each regularization method in Table \ref{tab:comp_cost}. Note that the calculation is based on a linear layer with weight $W_l \in \mathbb{R}^{d_o \times d_i}$ using a single sample. The computational cost has two sources: the estimation stage, where a small subset of the pre-training data is sampled to compute the FIM, and the training stage, where the regularization loss is computed at each iteration.

\begin{table}[htbp]
  \centering
  \caption{Comparison of computational cost and memory usage.}
    \begin{tabular}{lccc}
    \toprule
    \multirow{2}[4]{*}{\textbf{Method}\vspace{0.75em}} & \multicolumn{2}{c}{\textbf{Computation}} & \multirow{2}[4]{*}{\textbf{Memory}\vspace{0.75em}} \\
\cmidrule{2-3}          & \textbf{Estimation} & \textbf{Regularization} &  \\
    \midrule
    L2-SP & 0     & $\mathcal{O}(d_o d_i)$ & 0 \\
    EWC   & $\mathcal{O}(d_o d_i)$ & $\mathcal{O}(d_o d_i)$ & $\mathcal{O}(d_i d_o)$ \\
    KFAC & $\mathcal{O}(d_o^2 + d_i^2)$ & $\mathcal{O}(d_o d_i (d_o + d_i))$ & $\mathcal{O}(d_o^2+d_i^2)$ \\
    \bottomrule
    \end{tabular}%
  \label{tab:comp_cost}%
\end{table}%

\section{Experiments: Speech Synthesis}
\subsection{Tasks}
Having verified the efficacy of our methods quantitatively and objectively on language modeling tasks, we further apply them to our target application: the fine-tuning of speech synthesis models. Such models are typically more onerous and subjective to evaluate.  Our strategy is to demonstrate that the results from the objective evaluation also apply to the more specific target application.

Specifically, we fine-tune a pre-trained zero-shot speech synthesizer with LoRA to adapt it to an unseen speaker. Next, we evaluate the speaker similarity on both the target speaker and other out-of-domain (OOD) speakers, of which the former represents the fine-tuning performance and the latter indicates how well the model preserves the pre-training knowledge. To amplify the effect of catastrophic forgetting, the target speaker and other OOD speakers should be distinct from the pre-training data, thus we select speakers with particular accents for both fine-tuning and evaluation.

We appreciate that the task of evaluating the pre-training knowledge preservation is perhaps of less practical value since there is more interest in getting a similar voice to the target speaker than maintaining the zero-shot performance on other speakers in such a setting. However, this is a necessary compromise owing to several reasons.  Firstly, the current publicly available state-of-the-art speech synthesis models mainly target speaker adaptation and are far from being omnipotent, meaning a good zero-shot performance on other speech characteristics is not guaranteed.  Further, both the objective and subjective evaluation methods of speaker similarity are well-established, which is not the case for most of the others.  Finally, the multi-speaker speech data are easy to obtain, while in other cases the data are not. Despite the limitation, we believe the results will provide practical guidance not only for speaker adaptation on this model but also for many other models and usages where catastrophic forgetting is detrimental to the model's inherent capabilities.

\subsection{Model: StyleTTS 2}
To proceed with the proposed tasks, we need an open-access pre-trained TTS model that has good synthesis quality and zero-shot performance for speaker adaptation. StyleTTS 2 \cite{styletts2} is a recently proposed end-to-end TTS model that utilizes style diffusion and adversarial training with a large speech language model to generate human-level expressive and diverse speech. It also achieves a remarkable zero-shot performance though only trained on limited data of 245 hours from the LibriTTS dataset \cite{libritts} compared to large-scale models such as VALL-E \cite{valle}, which is trained on 60k hours of data. Initial experiments on zero-shot synthesis show that despite StyleTTS 2 rendering excellent synthesis quality, the synthesized speech tends to lose the accent traits of the target speaker, which can be attributed to the limited training data. Nevertheless, this could be suitable for our experiments as it makes the improvement brought by fine-tuning or the degradation of zero-shot performance more distinguishable.

StyleTTS 2 has a variety of components, many of which are composed of modules that are not compatible with LoRA or whose Hessian estimation needs extra calculation, such as LSTMs and 1D/2D convolutions. However, we found in our initial experiments that only fine-tuning the linear modules in StyleTTS 2 already achieves reasonably good performance. Therefore, for convenience, we only fine-tune the linear modules in all components that are useful for inference of StyleTTS 2.

\subsection{Experimental Details}
\paragraph{Implementation.} Our code is based on the official implementation of StyleTTS 2 \footnote{\url{https://github.com/yl4579/StyleTTS2}}. The same PEFT library for previous experiments is used for applying Bayesian methods and LoRA to the model.

\paragraph{Hessian estimation.} We use the official fine-tuning code to calculate the Hessian estimates, during which all training losses are enabled to ensure the gradients are properly back-propagated to all components. Based on the experience from language modeling experiments, we randomly sample 1,000 utterances from the \texttt{train-clean-360} subset of the LibriTTS dataset for Hessian estimation to ensure accuracy.

\paragraph{Data.} We select \texttt{p248}, a female speaker with an Indian accent in the VCTK dataset \cite{vctk} as the target speaker and randomly split the data into the training set of 356 utterances (approximately 21 minutes) and the test set of 20 utterances. For OOD speakers, we select another 9 speakers (5 females, 4 males) with different accents from VCTK and randomly choose 20 utterances of each speaker as test sets.

\paragraph{Training and inference.} We adopt the official multi-stage fine-tuning strategy of 50 epochs described in the code repository for all models, only reducing the batch size from 8 to 2 due to hardware limits. LoRA is applied to the linear modules in all components except for the discriminators and the text aligner which are fully trained and only used during training. The rank and the scaling factor of LoRA are set to 16 and 2 respectively, resulting in an overall percentage of trainable parameters of 1.639\% (2.26M of 138M). The fine-tuning is conducted 3 times with different random seeds. For inference, we synthesize test samples using the test sentences for every speaker using the fine-tuned model. All experiments were conducted on the same hardware as previous experiments.

\paragraph{Evaluation.} We conduct both objective and subjective evaluations, focusing exclusively on the speaker similarity. Essentially, we use the objective test results as the guideline for our experiments and corroborate our findings with subjective test results. More details are provided in the following sections.

\paragraph{Regularization.} Based on the fact that L2-SP is far inferior to other methods, we only experiment with EWC and KFAC in this section. The optimal regularization strength $\lambda$ is selected using the same criterion as in the language modeling experiments based on the results of the hyperparameter sweep. It is $10^3$ for both EWC and KFAC.

\begin{table*}[htbp]
    \centering
    \caption{Main objective test results of speech synthesis experiments.}
      \begin{tabular}{llC{6em}C{6em}C{6em}C{6em}C{6em}C{6em}}
      \toprule
      \multirow{2}[4]{*}{\textbf{Speaker}\vspace{0.75em}} & \multirow{2}[4]{*}{\textbf{Accent}\vspace{0.75em}} & \multicolumn{6}{c}{\textbf{Model}} \\
  \cmidrule{3-8}          &       & \textbf{Pre-trained} & \textbf{Full} & \textbf{Linear} & \textbf{LoRA} & \textbf{LoRA+EWC} & \textbf{LoRA+KFAC} \\
      \midrule
      \textbf{p248} (f, target) & Indian & 0.216  & 0.695  & 0.652  & 0.654  & 0.633  & 0.648  \\
      \midrule
      OOD All & -     & 0.293  & 0.159  & 0.204  & 0.203  & 0.224  & 0.280  \\
      OOD Female & -     & 0.325  & 0.184  & 0.226  & 0.227  & 0.247  & 0.291  \\
      OOD Male & -     & 0.254  & 0.127  & 0.175  & 0.174  & 0.196  & 0.267  \\
      \midrule
      \textbf{p225} (f) & English & 0.318  & 0.167  & 0.241  & 0.252  & 0.296  & 0.352  \\
      p234 (f) & Scottish & 0.385  & 0.221  & 0.257  & 0.240  & 0.274  & 0.297  \\
      \textbf{p261} (f) & Northern Irish & 0.448  & 0.206  & 0.288  & 0.281  & 0.323  & 0.374  \\
      p294 (f) & American & 0.267  & 0.131  & 0.173  & 0.181  & 0.166  & 0.241  \\
      p335 (f) & New Zealand & 0.205  & 0.195  & 0.171  & 0.179  & 0.176  & 0.188  \\
      \midrule
      \textbf{p245} (m) & Irish & 0.324  & 0.143  & 0.189  & 0.209  & 0.256  & 0.319  \\
      \textbf{p302} (m) & Canadian & 0.262  & 0.109  & 0.169  & 0.170  & 0.219  & 0.308  \\
      p326 (m) & Australian & 0.165  & 0.132  & 0.112  & 0.105  & 0.082  & 0.164  \\
      p347 (m) & South African & 0.262  & 0.123  & 0.232  & 0.210  & 0.228  & 0.276  \\
      \bottomrule
      \end{tabular}%
      
      \medskip
      \raggedright
      \hspace{2.2em}* A suffix (m/f) is added to the speaker name to indicate the gender. Speakers in bold are selected for subjective evaluation.

    \label{tab:tts_obj}%
  \end{table*}%

\subsection{Objective Evaluation}
For the objective evaluation, we use an ECAPA-TDNN \cite{ecapa-tdnn} speaker verification model \footnote{\url{https://huggingface.co/speechbrain/spkrec-ecapa-voxceleb}} to compute the averaged speaker embedding cosine similarity (SECS) score between the synthesized speech and the ground truth on the test set of each speaker. The averaged results of the three runs are shown in Table \ref{tab:tts_obj}. Note that OOD All/Female/Male are the aggregated scores of all/female/male OOD speakers, ``Full'' and ``Linear'' stand for full fine-tuning and linear module-only fine-tuning, respectively. We analyze the results from the following perspectives.

\paragraph{Fine-tuning performance.} After fine-tuning, the SECS score of the target speaker p248 increases from 0.216 to above 0.6, which manifests that fine-tuning is essential for improving speaker similarity. Without a doubt, the full fine-tuning achieves the best performance. The linear module only fine-tuning (``Linear'') and its LoRA-enabled counterpart (``LoRA'') perform similarly, however falling behind by a less than 10\% margin. This demonstrates the efficacy of the linear module-only fine-tuning scheme. Applying EWC and KFAC on top of LoRA further degrades the performance slightly, with KFAC performing slightly better than EWC. 

\paragraph{Zero-shot performance.} The overall scores on all OOD speakers clearly demonstrate the catastrophic forgetting, dropping from 0.293 for the pre-trained model to 0.159 for the fully fine-tuned model. Fine-tuning the linear modules only with or without LoRA slightly mitigates the forgetting, suggesting it is necessary to apply additional regularization. Under optimal $\lambda$ settings, KFAC (0.280) performs substantially better than EWC (0.224), only showing a slight degradation compared to the pre-trained model. The gender breakdown indicates that the fine-tuned model generally achieves a higher similarity on females than males, which can be attributed to the female fine-tuning data. This is confirmed by our test listening that the male speech synthesized by models without regularization severely deteriorates and resembles female speech more. In the speaker breakdown, despite the pre-trained model performing well on some speakers, the fine-tuning degrades similarities on all OOD speakers. One of the reasons for this could be the distinction between the target speaker and the OOD speakers in terms of the accent and the timbre. Moreover, the similarity drops more on speakers that previously had high similarity before fine-tuning. However, in any case, KFAC successfully preserves the zero-shot performance of the model, exceeding EWC by a large margin. 

\begin{table}[htbp]
    \centering
    \caption{Comparison of EWC and KFAC with varying regularization strength.}
      \begin{tabular}{ccccc}
      \toprule
      \multirow{2}[4]{*}{\textbf{$\lambda$}\vspace{0.6em}} & \multicolumn{2}{c}{\textbf{EWC}} & \multicolumn{2}{c}{\textbf{KFAC}} \\
  \cmidrule{2-5}          & \textbf{Target} & \textbf{OOD} & \textbf{Target} & \textbf{OOD} \\
      \midrule
      $10^{2}$ & 0.641  & 0.213  & 0.647  & 0.261  \\
      $\mathbf{10^{3}}$ & \textbf{0.633} & \textbf{0.224} & \textbf{0.648} & \textbf{0.280} \\
      $10^{4}$ & 0.575  & 0.270  & 0.593  & 0.283  \\
      $10^{5}$ & 0.379  & 0.271  & 0.491  & 0.271  \\
      \bottomrule
      \end{tabular}%
    \label{tab:tts_lambda}%
  \end{table}%

\paragraph{Regularization strength.} We provide the $\lambda$ sweep results in Table \ref{tab:tts_lambda}. It can be seen that under all $\lambda$ settings, KFAC always achieves better fine-tuning performance and better zero-shot performance preservation than EWC. When matching a good similarity score above 0.6 on the target, EWC shows a significant degradation on OOD speakers. Furthermore, as $\lambda$ increases, EWC's fine-tuning performance drops faster than KFAC and its zero-shot performance never surpasses that of KFAC. Overall, the results suggest that KFAC helps maintain the zero-shot synthesis ability of the pre-trained model while achieving good fine-tuning performance, whereas EWC suffers from a significant loss of fine-tuning performance when preserving the pre-training knowledge. This is consistent with the results of language modeling experiments on the smaller 350M model, however here the phenomenon is more pronounced.

\subsection{Subjective Evaluation}
\paragraph{Sample selection.} Having verified the efficacy with objective tests, we further conduct a subjective evaluation to corroborate our findings. One of the concerns is that the synthesized samples of OOD speakers usually result in a much lower perceptual similarity than those of the target speaker, making it difficult to distinguish the performance of low-performing models. In this regard, we select two OOD speakers that have the highest SECS scores and the most difference among models in each gender for the listening test, which are p225, p261, p245, and p302. 10 samples of the target speaker and 5 samples of each OOD speaker are randomly selected, totaling 10 female samples and 10 male samples of the OOD speakers for each model. We also add a ground truth (GT) group for comparison.

\paragraph{Implementation.} We hired 20 native English speakers from the United Kingdom on the Prolific \footnote{\url{https://www.prolific.com}} crowd-sourcing platform to rate the speaker similarity between the synthesized speech and the reference on a 5-point scale (5: completely same speaker, 4: mostly similar, 3: equally similar and dissimilar, 2: mostly dissimilar, 1: completely different speaker), using a modified Degradation Category Rating (DCR) method based on the P.808 toolkit \cite{p808}. The reference is a random recording of the speaker with spoken content different from that of the test sample and is bound to each test sample. The averaged result is often referred to as the Similarity Mean Opinion Score (SMOS).

\begin{table}[htbp]
    \centering
    \caption{Subjective test results with 95\% confidence interval.}
    \resizebox{\columnwidth}{!}{
      \begin{tabular}{lcccc}
      \toprule
      \textbf{Model} & \textbf{Target} & \textbf{OOD All} & \textbf{OOD Female} & \textbf{OOD Male} \\
      \midrule
      GT    & 4.46\,±\,0.11 & 4.59\,±\,0.07 & 4.65\,±\,0.10 & 4.52\,±\,0.11 \\
      \midrule
      Pre-trained & 1.90\,±\,0.15 & 2.22\,±\,0.13 & 2.36\,±\,0.20 & 2.08\,±\,0.17 \\
      \midrule
      Linear & 4.06\,±\,0.16 & 1.50\,±\,0.10 & 1.83\,±\,0.17 & 1.18\,±\,0.07 \\
      LoRA  & 3.86\,±\,0.16 & 1.48\,±\,0.09 & 1.83\,±\,0.17 & 1.13\,±\,0.06 \\
      LoRA+EWC   & 3.60\,±\,0.14 & 1.51\,±\,0.10 & 1.77\,±\,0.17 & 1.26\,±\,0.09 \\
      LoRA+KFAC  & 3.81\,±\,0.16 & 2.08\,±\,0.13 & 2.31\,±\,0.20 & 1.85\,±\,0.16 \\
      \bottomrule
      \end{tabular}%
    }
    \label{tab:tts_subj}%
\end{table}%

\paragraph{Results and analyses.} The results are shown in Table \ref{tab:tts_subj}. In general, the subjective test results corroborated our findings from objective tests, hence we mainly comment on the discrepancies between the two tests. For the target speaker, fine-tuning linear modules (``Linear'') achieves an SMOS of 4.06, which is a significant improvement from the pre-trained model of 1.90 and is considerably good given the ground truth of 4.46. Different from the objective test results, the LoRA-only model shows a disadvantage of 0.20 compared to ``Linear'', meaning fine-tuning a low-rank representation does degrade the fine-tuning performance for this model. The small difference between EWC and KFAC shown by SECS scores is actually perceivable, indicated by a difference of 0.21 in SMOS. In terms of zero-shot performance, EWC's preservation effect is not reflected on SMOS considering all OOD speakers, which is in contrast with KFAC. The gender breakdown shows a slight degradation on male OOD speakers for the LoRA with KFAC model, suggesting KFAC did not perfectly preserve the zero-shot performance of the pre-trained model as the SECS scores showed.

\section{Conclusions}
In this work, we explored applying Bayesian learning techniques to parameter-efficient fine-tuning to overcome catastrophic forgetting. We started from the derivation of the Bayesian transfer learning framework and demonstrated that PEFT could be regularized to preserve the pre-training knowledge as long as the parameter shift of the fine-tuned layers could be calculated differentiably. We then conducted experiments with LoRA on both language modeling and speech synthesis tasks to verify the efficacy of the proposed methods and compared the performance of different Laplace approximations. Our results show that catastrophic forgetting can be overcome by our methods without degrading the fine-tuning performance. Furthermore, the results on both tasks suggest using the Kronecker-factored approximations of the Hessian produces more effective preservation of the pre-training knowledge and better fine-tuning performance than the diagonal approximations, even though the former requires more data to be estimated accurately. 

Current limitations of this work include that it cannot be applied to PEFT techniques that add new components to the model such as bottleneck adapters; however this is not a serious concern given suitable techniques like LoRA already provide good fine-tuning performance.  Further, it is only feasible when at least part of the pre-training data is accessible.  Finally, the efficacy on larger (TTS) models has not been verified due to the inaccessibility to these models and hardware constraints. We would like to evaluate our methods on larger TTS models when they become publicly available in the future.

\section*{Acknowledgments}
This project received funding under NAST: Neural Architectures for Speech Technology, Swiss National Science Foundation grant \href{https://data.snf.ch/grants/grant/185010}{185010}.

\bibliographystyle{IEEEtran}
\bibliography{chl-ref-short}

\vfill

\end{document}